# Colossal magnetic phase transition asymmetry in mesoscale FeRh stripes

V. Uhlíř[1*], J. A. Arregi[1,2] and E. E. Fullerton[1]

[1] *Center for Memory and Recording Research, University of California, San Diego, La Jolla, California 92093-0401, USA*

[2] *CIC nanoGUNE Consolider, Tolosa Hiribidea 76, E-20018 Donostia–San Sebastián, Spain*

**Coupled order parameters in phase-transition materials can be controlled using various driving forces such as temperature, magnetic and electric field, strain, spin-polarized currents and optical pulses. Tuning the material properties to achieve efficient transitions would enable fast and low-power electronic devices. Here we show that the first-order metamagnetic phase transition in FeRh films becomes strongly asymmetric in mesoscale structures. In patterned FeRh stripes we observed pronounced supercooling and an avalanche-like abrupt transition from the ferromagnetic to the antiferromagnetic phase while the reverse transition remains nearly continuous over a broad temperature range. Although modest asymmetry signatures have been found in FeRh films, the effect is dramatically enhanced at the mesoscale. The asymmetry in the transitions is independent of applied magnetic fields and the activation volume of the antiferromagnetic phase is more than two orders of magnitude larger than typical magnetic heterogeneities observed in films. The collective behavior upon cooling results from the role of long-range ferromagnetic exchange correlations that become important at the mesoscale and should be a general property of first-order magnetic phase transitions.**

Understanding and ultimately controlling emergent phenomena at the mesoscale[1] requires quantifying the interactions and correlations of the individual constituents in complex materials[2-5] as well as engineered[6-8] or self-assembled systems[9,10]. Interaction of strongly correlated electrons and symmetry breaking in such materials often leads to ordered states such as charge ordering, superconductivity, ferromagnetism and multiferroicity featuring first-order phase transitions with coupled order parameters[11]. The nature of first-order phase transitions that exhibit an interplay between multiple degrees of freedom (*i.e.* electronic, structural and/or magnetic) is at the forefront of materials science. Examples include metal-insulator transitions (MIT) in oxides (*e.g.* manganites[12] and $VO_2$, ref. 13), the Verwey transition in $Fe_3O_4$ (ref. 14), and metamagnetic transitions from the antiferromagnetic order (AF) to ferromagnetic order (FM) in manganites[15,16], $CeFe_2$ alloys[17], $Mn_2Sb$ alloys[18], $Hf_{0.8}Ta_{0.2}Fe_2$ (ref. 19) and FeRh[20-24]. In such systems the nature of the transition can readily be tuned by external parameters such as strain[3,23,25], pressure[26,27], temperature, magnetic[23,28] and electric fields[29] and chemical doping[30,31].

First-order phase transitions are generally characterized by hysteresis and phase separation at the transition[2,32]. This phase inhomogeneity coupled with local disorder often results in a broad transition as local regions undergo the transitions at different temperatures or fields. One approach to studying the nature of the phase transition is to study materials at the scale of the inhomogeneity in the system, either using spatially sensitive probes or confining the material to the size of the heterogeneity. For confined systems the broad phase transition will often exhibit discrete jumps in the order parameter as discrete regions of the sample undergo the first-order transition[33,34]. For instance, the MIT in $VO_x$ and manganites is exhibited by the temperature-dependent resistance changes occurring in a series of steps in patterned nanostructures whereas a smooth transition is observed for films[35,36]. Transport measurements in a constricted geometry have further revealed considerable details about the nature of phase transitions and new emergent phenomena such as the second re-emergent transition observed in nanostructured manganite films[37].

Here we explore the heterogeneous nature of the magnetic phase transition in FeRh films patterned into mesoscale stripes. FeRh undergoes a first-order phase transition from an antiferromagnetic (AF) to a ferromagnetic (FM) phase (see Fig. 1a) upon heating from room temperature to above ~360 K in zero magnetic field[20,23]. This first-order transition exhibits a temperature hysteresis between heating and cooling cycles and the transition is accompanied by a volume increase of 1-2% (ref. 38), a reduction in resistivity[21] and a large change in entropy[39]. The magnetoresistance and magnetization changes enable new metallic memory cells[40], new approaches in magnetic recording[41] and magnetic refrigeration[5,39], and can be actively controlled with low power when coupled to a piezoelectric[25]. More fundamentally, this material has become a test-bed for exploring the interplay of structural, magnetic and electronic phase transitions in metallic systems including ultra-fast optically-induced phase transitions[42,43].

Figure 1b shows the temperature dependence of the net magnetic moment of a 25-nm-thick FeRh film grown epitaxially on an MgO (100) substrate together with the temperature dependence of the resistance of a 1.1-µm-wide stripe patterned from the same film. The onset of FM order with heating is seen as an increase of the magnetization and concomitant decrease in the resistance. The resistance and magnetization across the phase transition are continuous smooth functions of temperature (or magnetic field[23]) yielding a transition that is relatively broad (about 10K) as also seen in previous publications on FeRh films[23]. The broadening of the transition is linked to film heterogeneity originating from local structural variations where the first-order phase transition exhibits a phase coexistence of AF and FM domains with different transition temperatures as one progresses through the transition[44-46]. The coexistence of AF and FM domains has been imaged by a range of techniques revealing typical sub-micron AF and FM domain sizes[44-46].

We explored the nature of the phase transition by patterning the FeRh films into stripes (Fig. 1c) with lateral dimensions below 1 μm corresponding to the characteristic domain sizes typically seen in imaging[44-46]. We find that confining the lateral dimension allows us to track the phase transition in the individual domains via electrical transport experiments. When warming the sample from the AF to FM phase the width of the transition is similar to that observed for the film. However, since the domain sizes are comparable to the patterned sample size we observed a staircase-like transition (see inset of Fig. 1d for a 550-nm-wide stripe) as distinct regions undergo the metamagnetic transition. The resistance steps range from 0.1% to 3.0% of the total resistance change during the transition, corresponding to the FM activation areas of 0.002 $\mu m^2$ to 0.08 $\mu m^2$. The typical (median) step corresponds to regions of 0.005 $\mu m^2$. Surprisingly, we observe a quantitative change in the transition behavior upon cooling as seen in Fig. 1d. When cooling the sample the FM-to-AF transition shifts to lower temperatures than observed in the film and occurs in only a few discrete jumps. For the sample shown in Fig. 1d the majority of the sample switches at a single temperature indicating a collective response of the entire sample even though the sample size (2.2 $\mu m^2$) is much larger than the typical heterogeneous region observed in full film imaging. We argue below that this results from the fundamentally different interplay of magnetic correlations in the AF and FM phases with structural disorder.

To understand the origins of the enhanced asymmetry in the phase transition in FeRh stripes we have explored the response of the patterned FeRh magnetic stripes on multiple films, different substrates and various measurement protocols. The asymmetry shown in Fig. 1d is a general observation for stripes of a width on the order of 500 nm or narrower. However, the exact behavior can vary from sample to sample. Figure 2 shows resistance vs. temperature for a 50-nm-thick, 220-nm-wide and 2.6-μm-long stripe where the cooling transition occurs in a few steps instead of only one isolated transition as seen in Fig. 1d. However, the sizes of the FM-to-AF jumps upon cooling are much larger than those seen in the warming curve. The thermal cycling was repeated (Fig. 2) and demonstrates very high reproducibility for both the warming and cooling curves. This suggests that the heterogeneity is linked to the local microstructure and is not governed by random thermally-activated processes. Although the number of discrete jumps varies even for the stripes of identical size fabricated from the same film, the asymmetry in the properties with heating and cooling is consistently present. Therefore, the presence of structural heterogeneity does not, on its own, explain the qualitative difference between the AF-to-FM and FM-to-AF transitions.

We have further explored the role of the film structure on the asymmetric transition by comparing the results for FeRh films grown on MgO (100) and $Al_2O_3$ (0001) substrates[23]. The details of the structure of similarly grown films can be found in Ref. 23. These two substrates provide a different strain state of the FeRh film (Al$_2$O$_3$ induces tensile in-plane strain while MgO induces compressive strain). This strain shifts the transitions to higher temperatures on MgO and to lower temperatures on Al$_2$O$_3$. Furthermore, MgO (100) substrates yield FeRh (100) films while Al$_2$O$_3$ (0001) yield FeRh (111) that is twinned on the Al$_2$O$_3$ substrate. Despite this significant structural difference we observed a similar asymmetry in the transition when patterned into stripes (Supplementary Fig. 1) demonstrating that the asymmetry is not the result of strain relaxation when patterning[47] or specifics of the crystallographic orientation.

Other potential origins of the asymmetric transitions are the measurement current and dipolar magnetic fields. Because of the large difference in resistivity between AF and FM phases the measurement current could lead to correlations of phases due to local differences in Joule heating. However, we varied the measurement current by a factor of 5 (the power by 25) and did not see any significant change in the behavior (see Supplementary Fig. 2).

The dipolar magnetic fields arising from heterogeneous FM-AF phases are known to contribute to the width of the phase transitions[23]. The FM regions generate local dipolar fields on the order of hundreds of mT that together with an external magnetic field could alter the transition temperature of neighboring AF regions (as shown schematically in Fig. 3a). The phase transition temperature shift is roughly -9 K per 1 T of external magnetic field. Thus the fields generated by the FM phases can potentially perturb the transition of neighboring AF regions. For an in-plane external magnetic field, the dipolar fields generated by a FM region in a film can either add or subtract to the external field depending on whether the FM and AF regions are collinear or orthogonal to the external field. However, for the stripes with the external field aligned with the stripe axis the dipolar fields from the FM regions will mostly add to the external field, potentially triggering the phase transition of neighboring regions (Fig. 3a). Conversely, for out-of-plane external magnetic fields the dipolar fields from the FM region will oppose the external field and hinder the nucleation of the FM phase in neighboring AF regions. Similar arguments can be used to examine the internal demagnetizing fields within the FM grains during cooling. These effective fields could lead to correlated behavior that is qualitatively different for AF-to-FM and FM-to-AF transitions.

To test the role of dipolar fields we measure the response of a patterned stripe with magnetic fields applied either parallel to the stripe (longitudinal field) or perpendicular to the stripe and the sample plane (see Fig. 3b,c). In case of the longitudinal geometry we also varied the field magnitude (Fig. 3b). We observe a mean-field shift of the average transition temperatures dependent on the magnitude of the applied field as well as on the different directions of the applied field. However, the stripes show the same transition asymmetry even though the contribution of the dipolar fields from the FM regions acting on the AF regions has the opposite sign for longitudinal and perpendicular applied fields.

After eliminating other potential sources we argue that the colossal asymmetry observed on the mesoscale arises from the qualitative differences in the magnetic correlations in the FM regions compared to the AF regions. FM correlations are generally robust to local disorder[48] whereas most studies of AF systems show the magnetic correlations are shorter-range[49-53]. It is well known that long-range FM order persists even in granular or amorphous films[48]. In contrast, AF order is found to be sensitive to local structural disorder where the AF correlation lengths are limited by the crystalline correlations length. For instance, thin granular AF films such as those used for exchange biasing in recording heads exhibit a blocking temperature that results from thermal fluctuations of individual AF grains[49]. The behavior of these granular AF films is successfully modeled considering only the magnetic energy of individual grains and ignoring any magnetic correlations between AF grains. Even in epitaxial AF films the magnetic correlations, as measured by neutron scattering, are observed to be shorter or equal to the structural order[50,51]. Furthermore, in studies of ferrimagnetic $Fe_3O_4$ (100) films on MgO (100) there are a number of anomalous properties that are observed including high saturation field and superparamagnetic behavior that results from the presence of anti-phase boundaries (a shift of a half unit cell, ref. 52). The non-integer shift leads to frustrated magnetic order and limited magnetic correlations. Similarly, it is known from exchange bias studies that there is weak exchange coupling between the FM and AF phases compared to direct FM exchange, which has been studied in FeRh films[53].

The small resistivity steps we observe upon warming indicate that individual AF regions (of a typical area of 0.005 $\mu m^2$) undergo a first-order transition from AF to FM phases. This confirms that there is low correlation between AF regions[46], which makes the AF-to-FM transition in the FeRh stripes very similar to that observed for the full films. However, in the case of the FM-to-AF transition the ferromagnetic interactions stabilize the FM phase and the patterned stripes show larger supercooling compared to that observed in the full film[54]. When an AF region finally nucleates in the stripe the ferromagnetic exchange is broken and the AF phase will propagate through the stripe resulting in a sharp transition[55]. This behavior is not observed in the full films since there are always nucleation sites (most likely non-magnetic inclusions in the films or residual AF clusters) to initiate the FM-to-AF transition. In the case of mesoscale stripes the probability of finding one of these nucleation sites within the patterned area is reduced.

We can test this hypothesis by measuring minor loops such as was done in ferromagnetic films[56]. In Fig. 4 we compare the resistance vs. temperature loops where the AF-to-FM phase transition is not completed to the full thermal hysteresis. Stopping before a complete transition from the AF phase into the FM phase leaves residual AF domains in the stripe. Even stopping at 358 K where the AF-to-FM phase transition is 99.5% complete (as estimated by the change in resistance) results in a dramatic change of the FM-to-AF transition, which occurs at a much higher temperature and via multiple jumps. Warming to 354 K where the transition AF-to-FM transition is 97% complete results in the FM-to-AF transition occurring in many resistance steps and is qualitatively similar to that observed for a full film. This shows that only a few AF domains are enough to break the ferromagnetic exchange and nucleate the transition. When comparing films to mesoscale stripes it is always possible to find non-magnetic inclusions or defects that can nucleate the AF phase which will occur much less in the stripes. Notably, minor loops initiated on the cooling curve (see Supplementary Fig. 1), leaving residual FM domains in the AF phase before warming, do not significantly affect the nature of the AF-to-FM transition which is consistent with the limited exchange between neighboring AF domains and AF-FM domains[53] and lower magnetic correlation in the AF phase[46]. Furthermore, we see qualitatively similar behavior when cycling through the transition by magnetic fields at a fixed temperature (see Supplementary Fig. 3). This clearly indicates the asymmetry itself is not due to thermally activated processes.

The changes to the FM-to-AF phase transition is analogous to what has been observed in ferromagnetic films with perpendicular magnetic anisotropy[56]. Full films exhibit much smaller coercive fields than expected from the magnetic moment and anisotropy. Such behavior was historically known as Brown's paradox[57]. It is now well understood that this low coercive field arises because magnetization reversal occurs via nucleation of reverse domains at defect sites in the films followed by domain-wall propagation. The field to move a domain wall is much lower than the field needed to nucleate reversal in the absence of a defect. Patterning the film into mesoscale islands such that there are no longer any defect sites that nucleate reversal in the film within the area of the patterned region significantly enhances the coercive field. The magnetic reversal in single islands then proceeds via a single Barkhausen jump when a field-driven nucleation occurs[58]. The switching behavior of single islands is qualitatively different from the extended film even when the island size is large compared to the ferromagnetic domain size and exchange length.

In conclusion we have uncovered an unexpected asymmetry of the first-order AF-to-FM metamagnetic phase transition between warming and cooling in mesoscale FeRh stripes. While modest asymmetries have been observed in the studies of FeRh films[23,46,54], these effects are enhanced by orders of magnitude at the mesoscale and highlight the difference in magnetic correlations of the AF and FM phases. We believe these results should be a general behavior of first-order magnetic phase transitions of materials confined to the nano- and mesoscale. The nature of phase transitions initiated in AF or nonmagnetic phases should be much less sensitive to lateral confinement than the first-order transitions that initiate in the FM phase. This results from the robustness of the FM exchange to local strain and disorder when compared to the AF exchange. Similar behaviors are found in the non-magnetic MIT where strain can be a limiting factor[59]. These results will have to be considered for devices and applications utilizing FeRh[25,40,41] or related materials and may be potentially exploited in complex phase-transition materials, provided a transition specific order parameter can be effectively stabilized and controlled.

**Methods**

Epitaxial FeRh films were grown on MgO (100) substrates at 450 °C and an argon pressure of 1.5 mTorr by dc magnetron sputtering using an equiatomic target. The films were post-annealed at 800°C for 45 min and subsequently coated with a 2-nm Pt layer. The crystallographic orientation is such that [100] direction of FeRh aligns with [110] direction of MgO. The FeRh films were patterned into stripes by e-beam lithography and ion-beam etching. The Au/Ti transport leads were subsequently made by UV lithography combined with the lift-off technique.

The electrical transport measurements were performed using a Keithley SourceMeter 2400 in both the 4-probe and 2-probe configurations. However, as no qualitative difference in the resistance of the FeRh stripes upon temperature sweeping was observed, we used the 2-probe configuration to increase the number of measured samples. The probe current was 30 µA. The temperature was swept at a rate of 1 K/min and a single resistance measurement was taken every 0.5 s, *i.e.* the data are separated by a temperature step of 0.008 K. Unless stated otherwise, all measurements were performed in an applied field of 1 T.

**Acknowledgements**

The research was supported by the funding program of U.S. Department of Energy, Office of Basic Energy Sciences award #DE-SC0003678. J.A.A. acknowledges Ministerio de Economía y Competitividad (Spain) for travel support from award No. EEBB-I-2012-04226 (FPI exchange research visit subprogram).

**Author contributions**

V.U. and E.E.F. designed the experiment and wrote the manuscript which was commented on by all the authors. V.U. and J.A.A. fabricated the samples and carried out the transport measurements. V.U. analyzed the data and prepared the figures.

**Competing financial interests**

The authors declare no competing financial interests.

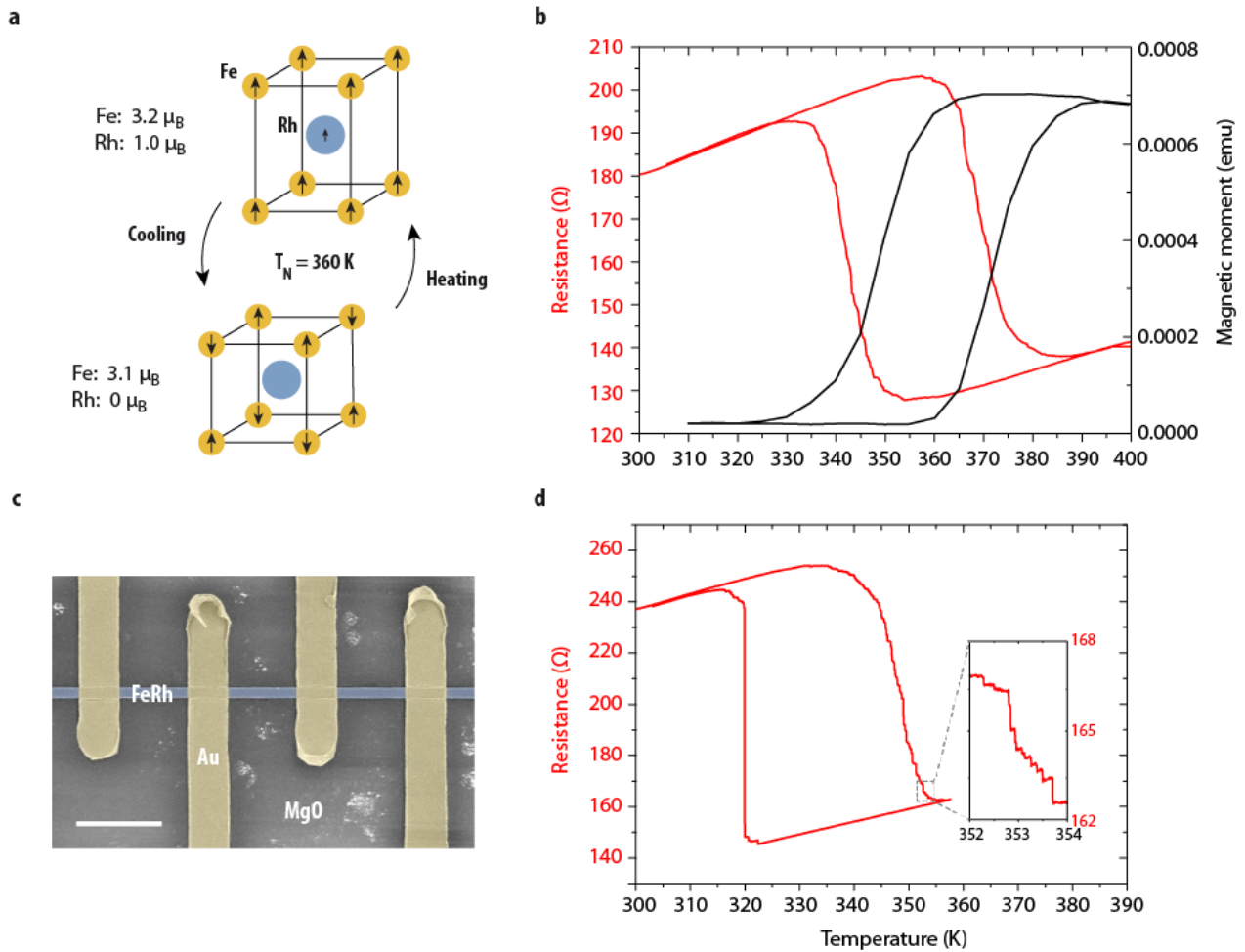

**Figure 1: Magnetostructural and transport properties of FeRh thin films and patterned stripes.**
**a,** The tetragonal cell of the AF FeRh film[23,24] at room temperature expands its volume when heated through the transition temperature to the FM state. The orientation of the AF-coupled planes was revealed by neutron diffraction[22]. The magnetic moments per atom[24] are indicated for each phase. **b,** Net magnetic moment vs. temperature of a 25-nm-thick FeRh film and resistance vs. temperature of a 1.1-µm wide stripe patterned from the same film. **c,** SEM image of the patterned FeRh stripe with transport leads. The scale bar is 5-µm-long. **d,** Asymmetry in the transition for a 35-nm-thick, 550-nm-wide and 4-µm-long stripe. Compared to full films and wide stripes, the phase transition is shifted to lower temperatures due to strain relaxation. The inset shows discrete steps in the order parameter upon heating corresponding to the transition in uncorrelated regions of the sample. Upon cooling the transition proceeds primarily through a single event.

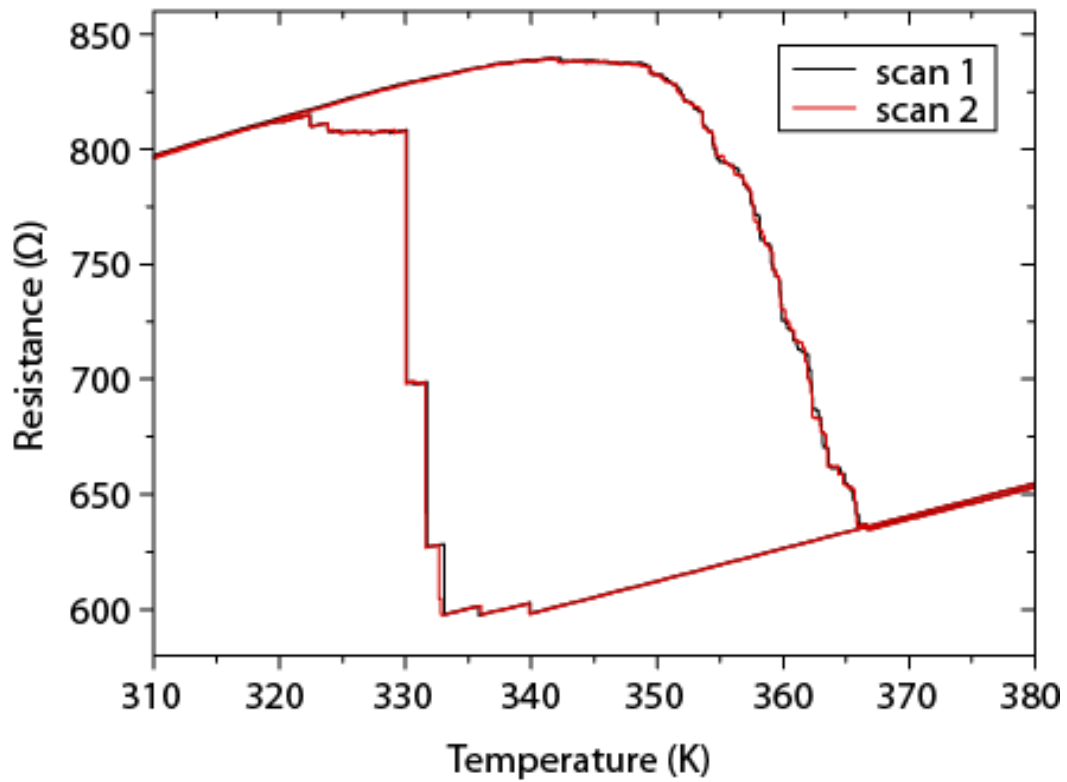

**Figure 2: Reproducibility of the resistance scans.** Dependence of resistance vs. temperature for a 50-nm-thick, 220-nm-wide and 2.6-µm-long stripe. The transition upon heating is continuous and uncorrelated, whereas upon cooling the transition occurs in a few abrupt steps. Two hysteresis loops are plotted showing an excellent reproducibility upon temperature sweeping.

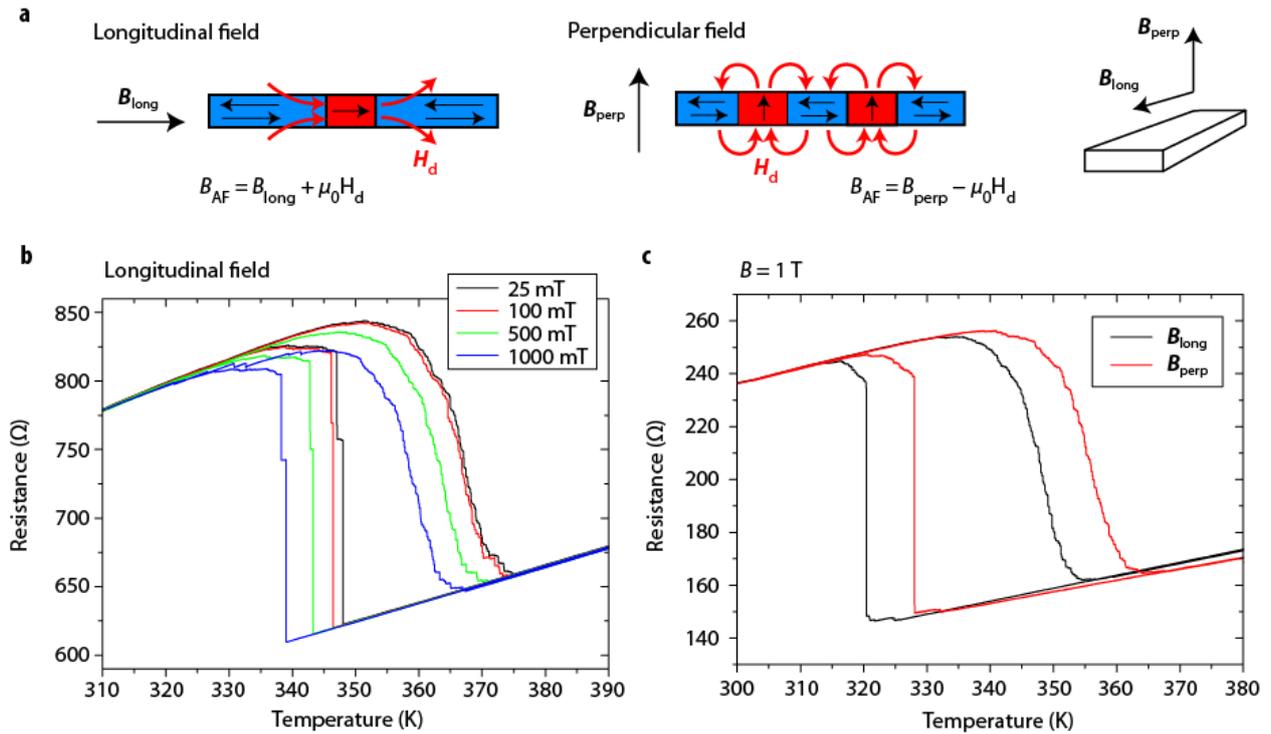

**Figure 3: Effect of the magnitude and direction of the applied field. a,** Applied magnetic field aligns the FM phase in the directions longitudinal and perpendicular to the stripe modifying the local effective fields acting on the AF phase due to different demagnetizing factors in the respective directions. **b,** Longitudinal magnetic field induces an offset of the thermal hysteresis loop of about 9 K/T, but no effect on the transition asymmetry is observed. **c,** In the direction perpendicular to the stripe the demagnetizing field almost entirely compensates the applied field (the offset is 7.6 K). The data shown in **b** are for a stripe 220-nm-wide and 50-nm-thick, c, a stripe 550-nm-wide and 33-nm-thick.

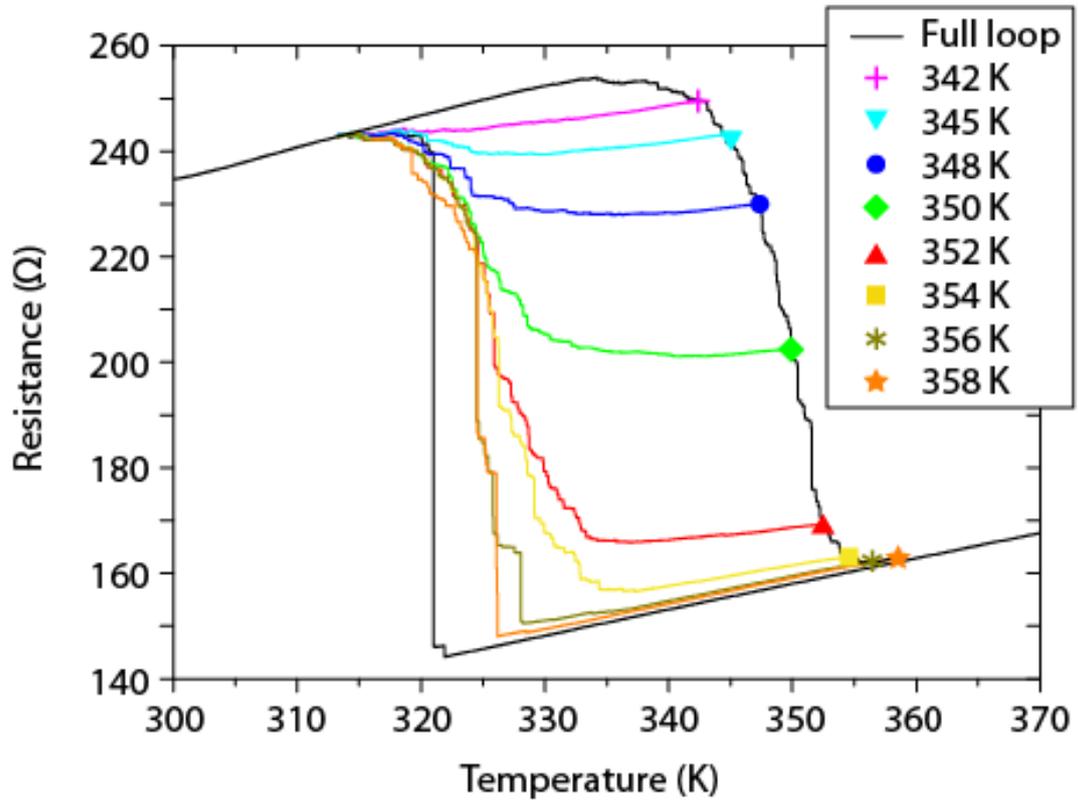

**Figure 4: Temperature-driven minor loops.** The black curve represents a major loop showing the full temperature hysteresis. Minor loops are initiated at different temperatures on the heating curve, indicated by symbols. The minor loop initiated at 358 K (marked by a star pentagon), where the AF-FM transition is 99.5% completed, shows reduced supercooling and multiple-step transition as the remaining AF clusters break the FM correlation.

# Supplementary information: Colossal magnetic phase transition asymmetry in mesoscale FeRh stripes

V. Uhlíř[1*], J. A. Arregi[1,2] and E. E. Fullerton[1]

[1] *Center for Memory and Recording Research, University of California, San Diego, La Jolla, California 92093-0401, USA*

[2] *CIC nanoGUNE Consolider, Tolosa Hiribidea 76, E-20018 Donostia–San Sebastián, Spain*

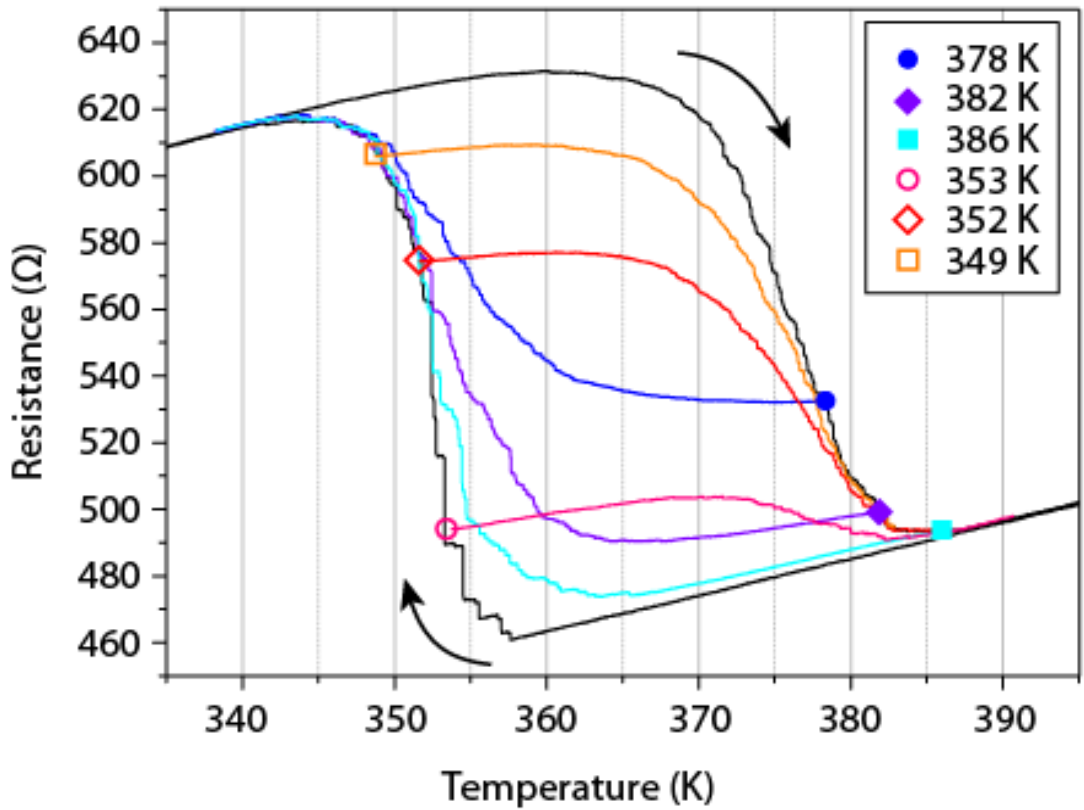

**Supplementary Figure 1: Resistance vs. temperature of an FeRh stripe on $Al_2O_3$ substrate.** Resistance vs. temperature of a 30-nm-thick, 400-nm-wide and 3.8-µm-long stripe. Multiple steps in the transition upon cooling indicate increased structural disorder due to (111) growth. The minor loops initiated upon heating (full symbols) and cooling (open symbols) are plotted. The minor loops initiated upon cooling do not show any change of the transition character upon warming with respect to the major loop. The minor loops initiated upon heating just before the completion of the AF-FM transition show a clear difference of the transition step size upon cooling with respect to the major loop – the long-distance FM correlation is broken by the AF residuals.

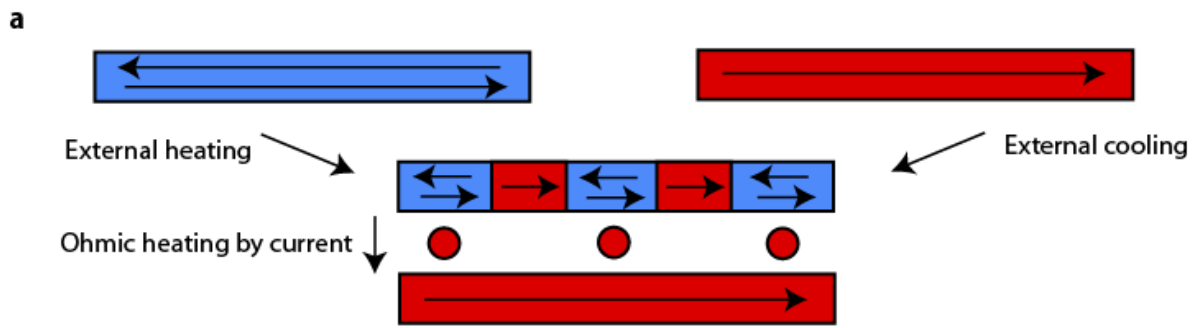

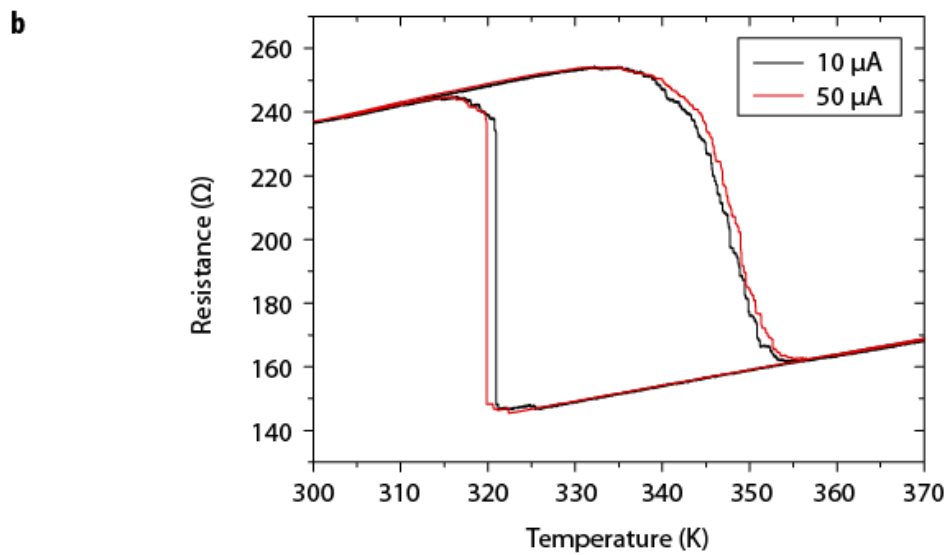

**Supplementary Figure 2: Effect of probe current. a,** The FeRh stripe is driven by temperature or magnetic field from either AF or FM phase into a phase-separated state. Once there, the probe current preferentially heats the AF phase due its higher resistance and thus stabilizes the FM phase. **b,** This effect did not influence the transition asymmetry while the probe current was varied by a factor of 5.

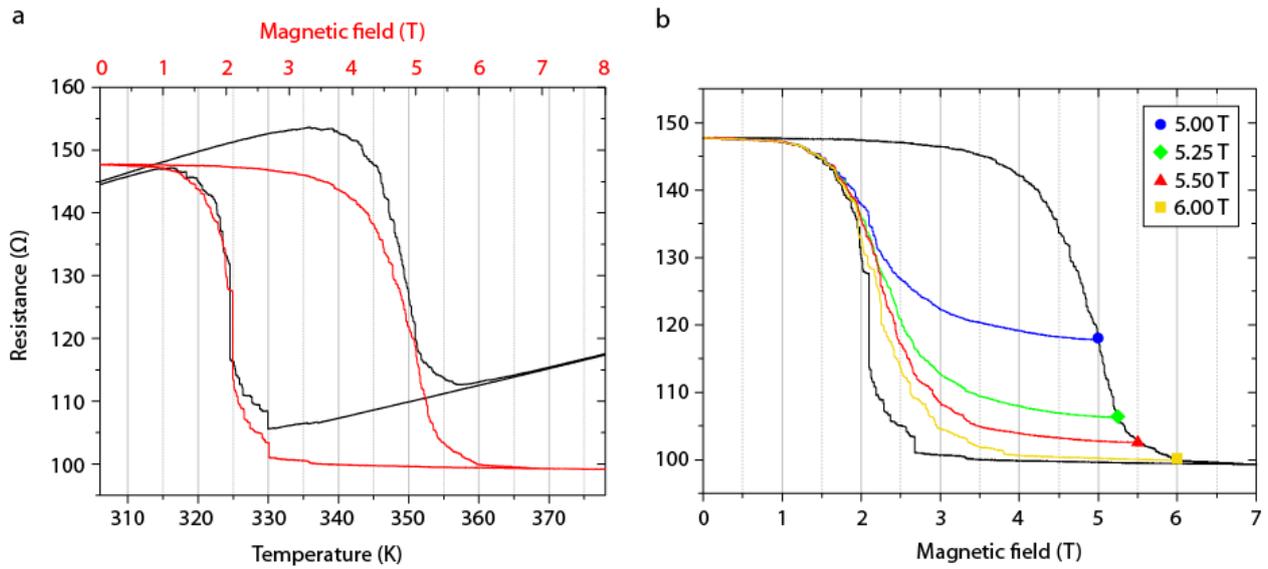

**Supplementary Figure 3: Field-driven magnetic phase transition. a,** Correspondence of the magnetic field- and temperature-driven phase transition hysteresis for a 33-nm-thick, 1.3-µm-wide and 4-µm-long stripe. The field hysteresis is taken at a temperature of 315 K, the temperature hysteresis at a field of 1 T. **b,** Minor loops initiated prior to completion of the AF-FM transition show qualitatively similar behavior as the temperature-driven ones: the AF residuals break the FM correlation and the transition becomes continuous.